# Inferring diffusion in single live cells at the single molecule level

Alex Robson[1], Kevin Burrage,[2,3] and Mark C. Leake[1,4,*]

[1]Clarendon Laboratory, Dept of Physics, Parks Road, Oxford University, Oxford OX1 3PU, UK, [2] Department of Computer Science, Wolfson Building, Parks Road, Oxford OX1 3QD, UK, [3]Mathematics Department, QUT, Brisbane QLD 4001, Australia; [4]Dept of Biochemistry South Parks Road Oxford, OX1 3QU, UK.

[*]Correspondence: m.leake1@physics.ox.ac.uk

**ABSTRACT**

The movement of molecules inside living cells is a fundamental feature of biological processes. The ability to both observe and analyse the details of molecular diffusion *in vivo* at the single molecule and single cell level can add significant insight into understanding molecular architectures of diffusing molecules and the nanoscale environment in which the molecules diffuse. The tool of choice for monitoring dynamic molecular localization in live cells is fluorescence microscopy, especially so combining total internal reflection fluorescence (TIRF) with the use of fluorescent protein (FP) reporters in offering exceptional imaging contrast for dynamic processes in the cell membrane under relatively physiological conditions compared to competing single molecule techniques. There exist several different complex modes of diffusion, and discriminating these from each other is challenging at the molecular level due to underlying stochastic behaviour. Analysis is traditionally performed using *mean square displacements* of tracked particles, however, this generally requires more data points than is typical for single FP tracks due to photophysical instability. Presented here is a novel approach allowing robust *Bayesian ranking of diffusion* processes (BARD) to discriminate multiple complex modes probabilistically. It is a computational approach which biologists can use to understand single molecule features in live cells.

**Keywords/phrases:** Diffusion, confinement, fluorescent proteins, in vivo imaging, single particle tracking, membrane heterogeneity

**1. INTRODUCTION**

Biological processes in the cell membrane are hard to replicate in artificial bio-mimetic membranes *in vitro* as the native protein-lipid architectures and dynamics in the membrane environment are far from well understood, even in the simplest prokaryotic organisms such as bacteria, let alone in more complex eukaryotic cells. An emerging paradigm for membrane sub-structure has changed from that of a freely-mixed system embodied by the classic Singer-Nicholson model [1] to the concept of a compartmentalized fluid [2-4]. It is the interactions between diffusing proteins and the underlying membrane sub-structure that maintains the observed heterogeneity. Several observations have led to this hypothesis; on a macroscopic length scale of several hundred nanometres, the diffusion coefficient of proteins are one to two orders of magnitude lower than those observed in artificial membranes [4-9], also the observation that membrane proteins have dramatic drops in diffusion rates upon oligomerization or aggregation [4, 10, 11], incommensurate with Saffman-Delbrück modelling [12, 13] which represents the standard analytical method for characterizing the frictional drag of protein molecules in lipid bilayers. Non-specific interactions are also attributed to membrane heterogeneity; for example, simple lipid bilayers protein-lipid and lipid-lipid interactions can cause proteins to partition into self-associating clusters [14], creating protein-rich or poor regions in cells. Also, there is some evidence for regions of lipid micro- and nanoscale structure identified in some eukaryotic membranes commonly referred to as lipid "rafts", which often appear to be consistent with mobile regions of phase-separated membrane that exist in an ordered, dense liquid phase surrounded by a more fluidic phase [15, 16]. These may be of functional advantage to signalling systems as well as being implicated in protein partitioning.

What is apparent is that there exists significant heterogeneity in local membrane architecture for a range of important biological functions. A key method for investigating the complex environment of the cell membrane is to monitor the fine details of diffusion of single molecules and complexes in native membranes. A tool of choice is fluorescence microscopy. This offers relatively minimal perturbation to native physiology whilst presenting an exceptional imaging contrast at single-molecule sensitivity levels that can allow the movement of individual fluorophore-tagged molecules, such as proteins and lipids, to be tracked with nanoscale precision [17-19].

Single particle tracking (SPT) approaches in general are powerful for interrogating dynamic membrane processes. Earlier studies involved colloidal gold for tracking [4, 20]. This has a clear advantage of an exceptionally high signal-to-noise ratio for particle detection with no danger of probe photobleaching, which permits longer tracks to be obtained with very short sampling time intervals at the sub-millisecond level. However, a significant disadvantage is the size of the probe at typically tens to hundreds of nanometres – this is often larger than underlying sub-structures of the membrane. SPT of fluorescently-labelled particles in the membrane offers significant advantages in using a much smaller probe on the nanometre scale. This was first applied using organic dye labelling [21, 22], but the recent use of genomically-encoded fluorescent protein (FP) reporters, such as green fluorescent protein (GFP) and its different coloured variants, has enabled many SPT studies to be performed on living cells with exceptional tagging specificity for the protein under investigation [23].

The most robust fluorescence imaging method for probing molecular level localization in the cell membrane is TIRF microscopy (see [24] for a discussion). This uses typically laser excitation at a highly oblique angle of incidence to generate an evanescent excitation field in the water-based environment of the sample - this can be thought of as an "optical slice" of ~100 nm thickness on the surface of the glass microscope slide/coverslip on which a cell sample is mounted. This results in significant excitation of fluorescently-labelled molecules in the cell membrane in the vicinity of the slide/coverslip surface. There is minimal excitation of components beyond this, either in the cell or from background fluorescence in the physiological buffer, therefore the signal-to-noise ratio for imaging membrane components is increased substantially. The end result is a very high detection contrast for fluorescently-labelled molecules and complexes in the cell membrane.

Four principal different diffusive modes are illustrated schematically in figure 1*a*, *b* with figure 1*c* plotting idealized mean square displacement (MSD) versus time interval *t* [25]. Brownian motion represents "normal" diffusion and is the simplest mode of diffusion characterized by a linear relation between MSD and *t*. However, a tracked protein trajectory for which the MSD reaches an asymptote at high *t* is indicative of confined diffusion, suggesting that the tracked protein is being trapped by its local environment - such corrals have been hypothesized as being important to forming nanoscale reaction chambers thereby greatly enhancing chemical efficiency [4, 9, 26-28]. Directed motion has an upwardly parabolic MSD function versus *t* and is seen for example during active diffusive processes such as those which occur when molecular motors walk on microtubules [29, 30]. The final class of diffusion has many examples in biology and is defined by so-called anomalous or sub-diffusive behaviour [31]. This motion is usually modelled as an MSD being proportional to $t^\alpha$ where $\alpha$ is a coefficient between 0 and 1; this mode may, for example, represent the percolation of a protein through the disordered media of the membrane, hopping across corrals or interactions with specialized domains [32-35].

What is apparent from the *in vivo* single particle tracking studies that have emerged over the past decade is that diffusion of molecules and complexes in living cells is in general not simple when viewed over the broad time scale of milliseconds to seconds relevant to many essential biological processes, and the reason for this may be fundamentally linked to critical sub-structural features of cells which are characterized over a length scale of a few to several hundred nanometres. There is therefore a compelling biological need to try to understand these complex diffusive processes.

Several analytical approaches have been attempted for characterizing diffusion in the cell membrane in addition to standard Saffman-Delbrück modelling, often involving heuristic approaches [11, 36]. A common approach has been to measure the ratio of the MSD with that expected from simple Brownian motion, embodied by a "relative deviation" parameter [26]. For directed motion, this parameter increases with $t$, whereas for confined motion, it decreases. However, these types of MSD analysis approaches are weak on several levels when applied to tracks generated from FP-tagged molecular complexes *in vivo*. Firstly, trajectories are determined from a low signal-to-noise ratio environment in which only tracks of short duration are able to be measured due primarily to the poor photophysics of the fluorophore resulting in scant, imprecise MSD information at high $t$ [37]. Secondly, trajectories are generated from a stochastic process, implying significant deviations from the idealized graphs of figure 1*c*. In addition, this method is highly reliant upon an accurate measurement of the diffusion coefficient, which in the noisy, heterogeneous environment of the cell membrane may prove very challenging.

An appeal of MSD analysis lies in its relative simplicity to address qualitative questions concerning the membrane environment, for example effects of molecular crowding or confinement [38, 39]. However, as figure 1*d, e* illustrate there is an expected statistical spread of the MSD traces for pure Brownian diffusion simply on the basis of diffusion being a stochastical process, that could be erroneously interpreted as a different diffusive behaviour judged by any individual MSD curve from a single particle trajectory. Population averaging can smooth out such variation to obtain average behaviour, but with the unfortunate result that we lose informative data concerning the biological heterogeneity of the ensemble of diffusing molecular complexes. Recent improvements to MSD analyses have involved applications of some diffusion propagators directly. The propagators define the probability distributions that a diffusing particle will be at a given distance from its origin after a given time. These methods have been used in estimating cumulative probability distributions to substantiate the presence of different non-Brownian diffusive modes [33, 40, 41].

Our new approach here is to present an inference scheme that can separate the distinct types of diffusive modes of individual trajectories without population averaging, and do so in a probabilistic fashion, given conditions imposed on real experimental data. This can then be combined with photophysical information to quantify molecular stoichometry of diffusing complexes thus allowing probing of non-trivial relations between the size of a molecular complex and how fast it moves *in vivo*. The inference of these diffusive modes is done using a Bayesian approach, incorporating *a priori* knowledge, based on both simulation and experiment. We denote this as <u>B</u>ayesian <u>r</u>anking of <u>d</u>iffusion (BARD).

Our study outlines the principles of the inference in light of the theory of diffusive processes. We describe the details of simulation using the different diffusive modes, and the inference algorithm used in separating out diffusive modes in a quantitative, probabilistic manner. We validate the inference using realistic simulated data, and apply to two different cell strains expressing FP fusion constructs to different membrane proteins. We obtain these live-cell data using TIRF microscopy. One cell strain expresses a single transmembrane helix probe in the cell membrane with a GFP fusion protein. The second strain is a yellow fluorescent protein (YFP) fusion to single twin-arginine translocation (Tat) protein complexes expressed in the cytoplasmic membranes of living bacteria, which exhibit significant real heterogeneity in terms of molecular stoichiometry, architecture and mobility.

The key concepts of these analyses are described in the Materials and Methods section of the main paper, with the specifics in the Supplementary Material.

## 2. MATERIALS AND METHODS

### (a) *Bacterial cell strain and preparation*

Two different *Escherichia coli* strains were used in our *in vivo* microscopy investigations. One was cell strain AyBC, as studied previously [11], using identical cell preparation conditions. This represents a heterogeneous, oligomeric membrane protein system. The cell strain contained a construct specifying a C-terminal enhanced YFP tag (Clontech Laboratories Inc., Mountain View, CA) to the native *E. coli* protein TatA on the cytoplasmic side of the membrane. The Tat system of bacteria translocates natively folded protein substrates across the cytoplasmic membrane through a nanopore whose walls are composed of subunits of the TatA protein (figure 2*a*). In addition to TatA, there are two other essential proteins in the Tat system, TatB and TatC, implicated both in substrate recruitment and gating of the TatA nanopore (figure 2*b*).

A second fusion construct was also investigated, denoted Helix1021-GFP (figure 2*b*). This represents a far less complex membrane protein system, which consisted of just a simple model membrane protein of a single membrane-spanning alpha-helix fused to a GFP tag on the cytoplasmic side of the membrane [42]. The fusion gene coding for this model membrane protein used the open reading frame *sll1021* in the cyanobacterium *Synechocystis sp.* PCC6803 as a start point, but expressing this as a membrane protein in *E. coli* for which there were no identified orthologues. The protein has an undetermined function but has been identified in the plasma membrane of *Synechocystis* [43], with the predicted gene product consisting of 673 amino-acids with a single predicted transmembrane alpha-helix close to the N-terminus. A portion of the *sll1021* sequence coding for 38 amino-acids including the predicted transmembrane alpha-helix was fused in-frame to the gene coding for GFPmut3* [44] with a linker of 5 asparagine residues. This construct was expressed in *E. coli* cells from the arabinose-inducible pBAD24 vector [45], with a predicted topology for the model fusion protein suggesting a 3 C-terminus residue overhang into the periplasm, 19 residues which then constitute the transmembrane alpha-helix spanning the cytoplasmic membrane, and 16 residues which overhang on the N-terminus side into the cytoplasm, followed by the GFP tag.

Cells of both strains were grown in Luria-Bertani (LB) medium [46] aerobically with shaking overnight at 37°C, and supplemented with 50 µg/ml ampicillin for correct antibiotic-resistant colony selection. Cells were diluted by 1:100 from the saturated cultured into M63 minimal media for sub-culturing, and were grown to mid-exponential phase typically for 3.5 hours at 30°C. For the Helix1021-GFP strain, L-arabinose was added to the culture at a final concentration of 2 mM. Cells were injected into a 5-10 µl flow-cell with poly-L-lysine-coated glass coverslips as the lower surface, allowed to settle for 10 min, washed with excess M63 and incubated with a 0.1% suspension of 202 nm diameter latex microspheres (Invitrogen Ltd., Paisley, UK) for 2 min to mark the coverslip surface, and washed with excess M63 buffer.

**(b) *TIRF microscopy and single particle tracking***

A home-built inverted TIRF microscope was used with either a 473 nm laser for GFP excitation, or a 532 nm excitation wavelength for YFP excitation, with excitation intensity in the range 250-500 W cm$^{-2}$ and measured depth of evanescent field penetration 110 ± 10 nm, with specifications as described previously [7, 11, 28, 47-50], using either 473 nm or 532 nm laser dichroic mirrors and notch-rejection filters (Semrock) as appropriate. The focal plane was set at 100 nm from the coverslip surface to image the cell membrane conjugated to the glass coverslip. Fluorescence emission was imaged at ~40 nm/pixel in frame-transfer mode at 25 Hz by a 128x128-pixel, cooled, back-thinned electron-multiplying charge-coupled device camera (iXon+ DV860-BI, Andor Technology). Images were sampled for typically ~8 s. Fluorescent particle positions on each time-stamped image frame were detected and fitted using automated custom-written image-analysis software which fitted a two-dimensional radial Gaussian function plus planar local background to the image intensity data for each candidate particle.

This generated the fluorescence intensity for each distinct diffusing "spot" of fluorescence in the cell of typical width 300-400 nm, either due to a TatA-YFP complex or an assemblage of Helix1021-GFP molecules, plus the local background intensity per pixel due to any autofluorescence and/or diffuse fluorescence components, and outputted the intensity centroid to a sub-pixel precision of ~40 nm for single FP molecules, and down to 5-10 nm for molecular complexes/assemblages containing more typically ~tens of FP molecules.

Tracks were generated from each particle provided tolerance criteria in subsequent image frames were satisfied on the basis of size, intensity and position of detected particles in subsequent image frames, for at least five consecutive image frames. The MSD versus time interval relation was then calculated for each particle trajectory, as described previously [11]. Using a Fourier spectral approach we were able to estimate the stoichiometry of these complexes through step-wise photobleaching of the relevant fluorescent protein molecule [7].

**(c)** *Implementing and validating the BARD algorithm*

*Generation of synthetic tracks for validation.* Two-dimensional simulated tracks for use in validation were generated in a standard way by a stochastic random walk process in MATLAB (The MathWorks, Natick, MA) to approximate real diffusion for the fluorescently-labelled proteins in cytoplasmic membranes of *E. coli* cells, sampling at the same 40 ms video-rate time interval as for experimental imaging, with track durations of typically 0.8 s (see figure S1, Supplementary Material).

*Bayesian formulation.* The general principle of Bayesian inference is to quantify the present state of knowledge and refine this on the basis of new data, under-pinned by *Bayes' theorem*, emerging from the definition of *conditional* probabilities (further details, see Supplementary Material). In words this is simply:

Posterior=(Likelihood x Prior)/Evidence

There are two stages in our statistical inference; parameter inference and model selection. Both use an application of Bayes' Theorem. The first stage infers the posterior distributions about each model parameter, which is defined as:

$$P(w|d,M) = \frac{P(d|w,M)P(w|M)}{P(d|M)}.$$

Here, *M* is a specific diffusion model, *w* is a model parameter and *d* represents SPT data, and a phrase "*P(A|B)*" means "the probability of *A* occurring given that *B* has occurred". This stage is independent of other models, but is conditioned on one

single model, *M*. Both the posterior and likelihood are conditioned upon the data *d*. We can now explain the three names of the terms above:

- The *likelihood*, *P(d|w,M)*: the probability distribution of the data for a given parameter.
- The *prior, P(w|M)*: the initial distribution prior to any conditioning by the data. Priors embody our initial estimate of the system, such as distribution of the parameters or the expected order of magnitude.
- The *posterior, P(w|d,M)*: the distribution of the parameter following the conditioning by the data.

The second stage in our statistical inference is model selection. This invokes another application of Bayes' Theorem:

$$P(M|d) = \frac{P(d|M)P(M)}{P(d)}.$$

*P(M|d)* is a number which is the model posterior, or probability. *P(M)* is a number which is the model prior, P(*d|M*) is a number which is the model likelihood and *P(d)* is a number which is a normalising factor which accounts for all possible models. This now generates the posterior (i.e. probability) for a specific model.

Linking the two stages in our statistical inference is the term *P(d|M)*, the model likelihood. This is also the normalisation term in the first stage. As model priors are usually flat (i.e. all models are expected equally), *P(d|M)* is often referred to as the "evidence", a portable unitless quantity. In the general case, comparing the *P(d|M)* values for each independent model allows us to rank and select models (Supplementary Material)

*Diffusion Models*. As proof-of-principle, we used four standard diffusion models which are typical of observed molecular scale motion in living cells (Tables S1 and S2, Supplementary Material). These were Brownian, anomalous, confined and directed diffusion, and we used the underlying propagators associated with each different diffusion model directly (full details in Supplementary Material).

*Inference in BARD*. The inference scheme was split into two forms. One uses the likelihoods based on the mean square displacement distribution of each track, which we call the MSD method. The second uses the probability distribution functions directly on the individual frame-by-frame spatial displacements measured for each track, which we call the PDF method. These form the likelihoods, $P(w|d,M)$ (Supplementary Material).

As discussed in the Results section, the PDF method performed more accurately in many applications for comparing just two different non-confined diffusion models, such as anomalous diffusion with Brownian, but could not be applied to cases of confined diffusion, in which circumstance the MSD method was applied. Both approaches result in an estimate for the preliminary likelihood associated with each given single particle track.

The prior distribution for the diffusion coefficients $D$ for Brownian diffusion, and the equivalent transport coefficient $K_\alpha$ for anomalous diffusion, were modelled as Gamma distributions (Supplementary Material, and see ref. [51] for a discussion of using a Gamma distribution). The prior distributions for the effective characteristic confinement radius $R$ for the confined diffusion model, and for the mean drift speed $v$ for the directed diffusion model, were both approximated as exponential distributions with expected sizes in the range of values that had been measured from several earlier studies in other biological systems (see Table S3, Supplementary Material). The $\alpha$ factor in the anomalous diffusion model was assumed to be uniform (i.e. flat) in the range 0.5-1.0 without further modelling. We have no *a priori* expectations to indicate how this factor would be distributed. The literature at present suggests multiple models of sub-diffusion, so the sensible consensus prior in light of this would be flat. However, an extension would be to discriminate between these different families. Either way, our uniform assumption can account for the experimental observations of anomalous diffusion with an anomalous coefficient of ~0.7-0.8.

*BARD Implementation.* To implement our BARD algorithm, the following steps were taken:

1. Quantify all of the microscopic diffusion coefficients, $D_m$, from the single particle tracking data (shown here for simulated Brownian diffusion tracks in figure 4a). Here, $D_m$ gives a measure of the short time scale rate of diffusion and is estimated from a linear fit of the MSD data of each individual track using the first four data points (full details in Supplementary Material).

2. Fit a Gamma distribution to the distribution of all $D_m$ (figure 4*b*) and use this fit to generate the two characteristic shape parameters of this function. Then use these shape parameters to generate the diffusion coefficient prior (see Equation S11 and S12, Supplementary Material, and figure 4*c*, top panel).
3. Calculate the other parameter priors for $\alpha$, $R$ and $v$ for the anomalous, confined and directed diffusion models (Table S3, Supplementary Material).
4. For each separate single particle track we then calculated the *likelihood* (either using Equation S8 for the PDF method, or Equation S9 for the MSD method, see Supplementary Material).
5. We then estimated the unnormalized posterior for each single particle track against each diffusion model, taken for the pure Brownian diffusion model as:

    Posterior = Likelihood(Brownian propagator) x Prior($D_m$)

    For the anomalous diffusion model as:

    Posterior = Likelihood(anomalous propagator) x Prior($K_\alpha$) x Prior($\alpha$)

    For the confined diffusion model as:

    Posterior = Likelihood(confined propagator) x Prior($D_m$) x Prior($R$)

    And for the directed diffusion model as:

    Posterior = Likelihood(directed propagator) x Prior($D_m$) x Prior($v$)

    An example of the unnormalized Brownian model posterior distribution for a typical simulated track is shown in figure 4*c*, lower panel. The posteriors for the other diffusion models are shown for the same example track in figure 4*d-f*.
6. Normalize the parameter posterior distributions from stage 4 (details in Supplementary Material, equation S4), calculating the evidence term. This is the final step in the parameter inference section, which bridges to the second inference stage (i.e. model selection).
7. Model selection: Calculate the model posterior. Rank the models on the basis of the size of the model posterior (a numeric probability). This final step then yields a probability estimate for a given model, relative to all the other models investigated: $P(M|d)$. For example, for the example track shown in figure 4a, which was simulated using a pure Brownian diffusion propagator function, the inference ranking probabilities which were generated from the four candidate diffusion models of anomalous, Brownian, directed and confined are 33.1%, 65.6%, 1.1% and 0.2% respectively, and so in this instance Brownian diffusion is the favoured model. This is not to say that the absolute probability that the Brownian diffusion model is the correct one is ~66%, but rather that it has the highest probability of being true from the set of candidate models investigated.

8. For the top-ranked diffusion model for each single particle track we then automatically locate the centroid of the posterior, to indicate the specific value of the transport parameter for that particular diffusion model. This is done using a Gaussian fit about the posterior peak.
9. Repeat this process for all single particle tracks in the data set.

*Modelling mobility changes due to switches in diffusion coefficient.*

In order to demonstrate that the framework presented here can be extended to even more complicated cases of heterogeneous diffusion environments, we simulated a change in lateral diffusion coefficient as might be experienced by a single molecular complex undergoing transitions to multiple kinetic states. This may occur in signalling systems with transitions between ligand-bound and unbound states, or be due to a change in lateral mobility due to interactions with the underlying membrane such as local changes in viscosity [16] or interactions with the membrane cytoskeleton.

## 3. RESULTS

### (a) *TIRF microscopy on live bacterial cells*

Bespoke video-rate TIRF microscopy at 40 ms per frame (figure 3a) was performed on GFP-labelled Helix1021 and YFP-labelled TatA membrane protein complexes, resulting in the appearance of multiple distinct diffusing fluorescent "spots" in each cell that could be tracked automatically from frame to frame. These spots were typically ~300-400 nm in width. This was larger than we measured for the point spread function width from single fluorescent protein molecules immobilized to the surface of the coverslip by ~100 nm [7]. The measured point spread function width of single FP molecules of ~200-300 nm is equivalent to the optical resolution limit of our microscope and is an inevitable feature due to diffraction of emitted fluorescence when the detector, in our case an EMCCD camera, is physically more than a few wavelengths distance away.

The TatA system had been characterized previously using epifluorescence microscopy that indicated multiple spots per cell (mean of ~15) with a range of fluorescence intensities, diffusing over the cytoplasmic membrane surface [11]. Our aim in the present study was to use TIRF illumination to improve the imaging contrast sufficiently to generate single particle trajectories in the TIRF evanescent field in the specimen focal plane, corresponding to localization of either the Helix1021 or TatA in the cytoplasmic membrane. This would then permit analysis of the transport properties of these proteins at the single molecule/single molecular complex level for a relatively simple membrane protein probe at one extreme and for a complex heterogeneous membrane protein molecular complex at the other, both in functional, living cells.

Using automated single particle tracking [11] we were able to track individual fluorescent spots to a super-resolution precision of ~40 nm or less. Experimental single particle tracks were collated and MSD values estimated (full details in Supplementary Materials). The longest duration tracks lasted typically ~1 s, but in most cases the tracks were shorter, with ~10 data points per track being more typical.

For the TatA-YFP data, cells contained typically ~2-3 fluorescent spots in TIRF images (figure 3*b*), suggesting ~12-18 spots per cell since the TIRF evanescent field of our microscope we estimate encapsulates roughly 1/6 of the *E. coli* cell membrane. Most MSD traces indicated putative evidence for Brownian diffusion, with typical values of $D \approx 0.01~\mu m^2\,s^{-1}$. This was consistent with the earlier investigation, but with qualitative evidence from some putatively asymptotic MSD traces for a smaller sub-population of relatively immobile spots, as had been reported in the previous study but not robustly quantified [11]. Prior, likelihood and posterior distributions were estimated for each single particle track.

The Helix1021-GFP cells contained typically ~4-6 fluorescent spots per TIRF image (figure 3*c*), suggesting more like ~30 spots in total in the whole cell membrane. MSD data again indicated putative evidence for two populations in terms of diffusive modes, one of Brownian diffusion with typical values of diffusion coefficient higher by factor of ~5-10 than the TatA-YFP data, and the other mode again qualitatively suggesting confined diffusion.

**(b)** *Model ranking and parameter estimation*

To validate our approach we tested the inference method using realistic simulated two-dimensional SPT input data utilizing mobility parameters with characteristic values comparable to those estimated qualitatively for the experimental Helix1021-GFP and TatA-YFP data from the MSD plots. We then analyzed both the parameter estimates and model rankings outputs. The correctness of the model ranking was assessed by *classification matrices*. A classification matrix represents different "input" simulated diffusion models down the rows, *i*, while the different "output" diffusion models from the ranking inference are represented across the columns, *j*, and then each location in the matrix is given an associated number for the percentage of tracks that are included in that particular (*i*, *j*) class combination.

In figure 5 we show the results of two example classification matrices, one corresponding to likelihood estimation using the MSD method in figure 5*a*, the other to the PDF method in figure 5*b*. Previous experiments on other biological systems which involve examples of directed diffusion, for example with putative protein treadmilling studies *in vivo* [29], suggested different values for characteristic diffusion coefficients, and so directed diffusion model ranking was done separately (Supplementary Material). In this case, we investigated a range of different drift speeds from 1-20 nm s$^{-1}$. This indicated that at typical drift speeds used the directed diffusion model can be correctly identified against a Brownian model with a relative probability of ~60-70%.

Using the MSD method, confined and Brownian diffusion could clearly be identified correctly with greater than 50% relative accuracy, though the true identification of anomalous behaviour was poorer (~30%), probably because subtle sub-diffusive behaviour is not apparent for such typically short track lengths of only 10-20 data points as used here. The inference output for confined diffusion in particular was unsurprisingly found to be a function of track length, with the 50% threshold of correct inference for tracks being composed of at least ~16 data points (figure S2, Supplementary Material), though the change in correct relative inference probability for confined diffusion was found to be only a few % when the confinement radius was varied across a relatively large range 50-200 nm in estimating the posterior distribution. This is not to say that the choice in prior function has little effect on the final outcome; if we use a naïve "flat" prior function for the confined diffusion model (in effect, taking an infinitely large value for the confinement radius) then we estimate that the correct relative inference probability is over 20% lower compared against the non-flat priors used. In other words, utilizing physically sensible prior functions makes a substantial difference to correctly inferring the underlying type of diffusion (see Supplementary Material)

The PDF method is an approach which utilises information from the *relative* displacements of a tracked molecule or complex from frame to frame, and so can not be applied to a confined diffusion model without *a priori* knowledge of the *absolute* position of the diffusing particle relative to the boundaries of the putative confinement zone, which in general is not the case. Therefore, for the PDF method we display in figure 5*b* the relevant classification matrix between just anomalous and Brownian diffusion models. Here, anomalous diffusion was correctly discriminated with an accuracy of at least 62%, performing better than the MSD method for corresponding diffusion models (for example, Brownian diffusion was correctly identified with a relative accuracy of 95% using the PDF method compared with 52% for the MSD method).

**(c)** *Identifying switches in molecular mobility*

Tracks were simulated to mimic a sudden change in effective lateral diffusion coefficient, which have been observed previously in biological systems where multiple diffusion states exist [56]. Such multiple diffusion states might, for example, be due to either a dramatic change in lipid viscosity for the micro- or nanoscale environment in which a protein molecule or complex is diffusing, or conversely through a rapid oligomerization or molecular assembly process of the diffusing complex. In this simple generalization, we assumed that the time scale of the transitional step between different lateral diffusion coefficients was much less than the sampling period. For simplicity, we assumed that diffusing particles make this mobility switch at the halfway point of their full simulated trajectory. At this point, particles were assumed to switch to a higher diffusion coefficient (from 0.01 $\mu m^2 s^{-1}$ to either 0.05 $\mu m^2 s^{-1}$ or 0.10 $\mu m^2 s^{-1}$), assuming true Brownian diffusion in each case and a video-rate sampling time interval of 40 ms for which the number of data points in each half of a trajectory is $N = 10$.

A *switching* inference model was formulated by separating the displacement data at each time point and allowing for two separate mobility measurements to be inferred either side of this. Figure 6 illustrates the typical simulated individual and time-averaged MSD outputs with model ranking predictions. This relatively simple switching inference modification can correctly predict switching behaviour characterized by two separate microscopic diffusion coefficients over a simple Brownian diffusion mode characterized by just a single microscopic diffusion coefficient, with a relative ranking probability in the range 65-85%, depending upon the size of the switch in diffusion coefficient. Using the PDF approach under the same conditions generated a slight improvement to correct identification, and in doing so we found that the correct switching model was identified in preference to simple Brownian motion (that is, a ranking probability in excess of 50%) down to as small a change as ~3-fold in the microscopic diffusion coefficient.

**(d)** *Application of BARD to live-cell experimental data*

Preliminary inspection of the MSD traces generated from automated single particle tracking from both the Helix1021-GFP and the TatA-YFP *E. coli* cell strains suggested a predominantly mobile population with roughly linear MSD versus time interval traces, in addition to a relatively immobile population characterized by putatively asymptotic MSD versus time interval traces, which could be indicative of two possible populations corresponding predominantly to Brownian diffusion and confined diffusion. In the first instance, we ran a BARD analysis using all four standard diffusion models of anomalous, Brownian, confined and directed diffusion, which clearly indicated for both cell strains that Brownian and confined were the two most inferred diffusion models. We then pooled the combined inferred results from anomalous, Brownian and directed diffusion as constituting "mobile" tracks, and compared this to the inferred confined track data on MSD versus time interval plots.

Simulated realistic track data using our standard set of mobility parameters (Table S3, Supplementary Material) indicated that a mixture of such mobile and confined tracks could be successfully discriminated, with both the imposed values for microscopic diffusion coefficient and confinement radius agreeing with those inferred from the BARD analysis to within the measurement error (Figure 7*a*).

Applying BARD analysis to the Helix1021-GFP track data indicated that 50-60% of all tracks exhibited confined diffusion with an estimated confinement radius of 110 ± 50 nm (± s.d.), with the mobile population characterized by a microscopic diffusion coefficient typically in the range 0.01-0.05 $\mu m^2 s^{-1}$ (Figure 7*b*). BARD analysis applied to the TatA-YFP track data indicated a smaller but still significant proportion of 30-40% of all tracks exhibiting confined diffusion with a mean confinement radius of 60 ± 40 nm, and the mobile population characterized by a smaller typical microscopic diffusion coefficient in the range 0.002-0.01 $\mu m^2 s^{-1}$ (Figure 7*c*).

For both the Helix1021-GFP data (AR and MCL, manuscript in preparation) and the TatA-YFP data (Figure 8) we were able to estimate the molecular stoichiometry of the diffusing fluorescent spots using a Fourier spectral technique that utilized the step-wise photobleaching of fluorescent proteins [7]. This indicated a difference between the mobile and confined track populations suggesting that confined tracks were associated a greater typical number of fluorescent protein subunits. For example, we estimated the median stoichiometry from the mobile TatA-YFP spot population as being in the range ~20 TatA-YFP molecules per spot, whereas that of the confined population was higher by ~50% (figure 8*a*). We saw no obvious differences in microscopic diffusion coefficient between the confined and mobile populations (figure 8*b)* nor of any clear correlation between molecular stoichiometry in each fluorescent spot and the inferred size of the confinement radius (figure 8*c)*.

## 4. DISCUSSION

The ability to monitor single molecules or complexes diffusing in living cells is an excellent example of the "next generation" single-molecule cellular biophysics approaches which have emerged over the past decade. What some researchers are now trying to do with such exceptionally precise molecular-level data is to use them to increase our understanding of the functional architecture of both the diffusing molecules themselves and of their local cellular environment. However, to do so requires a development of novel computational methods that can accurately measure the underlying modes of diffusion from the typically noisy and limited data from these tracked molecules *in vivo*.

In this study, we describe a novel analytical method to discriminate different modes of diffusion, applicable to data obtained from single particle tracking of fluorescently labelled proteins in the cell membrane. Although the bio-computational algorithm in itself is complex it should find broad application for researchers in the cell biology field. We report our approach based on both modelling and stochastic simulation of multiple biologically-relevant diffusive modes experienced by proteins in different underlying micro- and nanoscale environments. Priors are formulated from both simulation and experimental work. We demonstrate how the use of the correct propagator functions can permit discrimination between Brownian, directed, confined and anomalous diffusion, even for relatively sparse data tracks. When comparing two diffusive modes in a pair-wise fashion our results indicate that model ranking predicts the correct diffusive mode for a single video-rate sampled track as short as ~0.4 s in duration. Furthermore, the model can be extended to permit discrimination for a diffusion model involving sudden switching of the diffusion coefficient during a particle's trajectory.

Two approaches were investigated, one using the MSD and the other using the PDF method. In each case, a prior formulation was used to describe the expected distribution of the parameters. Although neither approach could effectively discriminate between anomalous and confined modes of motion, which from the MSD curves have qualitatively similar shapes for noisy short tracks, we find that the PDF and MSD methods in tandem have different resolving power. The MSD method effectively identifies confined from simple Brownian motion, whereas the PDF method effectively identifies anomalous diffusion from Brownian diffusion. In addition, the PDF approach has a strong resolving power in that it can identify dynamics within a single track, as observed in the simulations of diffusion coefficient switching.

The PDF method does not take into account the full effect of experimental noise, as distinct from random fluctuations due to the stochastic nature of the diffusion processes. Levels of experimental noise are likely to vary between different experimental equipment and need to be properly characterized for each individual case. However, this was qualitatively incorporated into the MSD approach, where Gaussian errors are assumed. Experimental noise, arising in tracking would be included in the error, and would add by quadrature to the expected fluctuations due to stochastic noise (the time interval zero point in our case assumes an MSD error of around 40 $nm^2$).

Our *in vivo* video-rate particle trajectories contain approximately 10-30 times fewer data points than those used in previous studies utilizing tracking of gold particles [4, 26, 27, 53], organic dye labelling of clusters containing hundreds of molecules [39] or quantum-dot tracking [41], and are of comparable duration to those obtained previously using single molecule fluorescence microscopy either in artificial lipid layers or *in vivo* [11, 33, 40]. These have implemented a variety of different methodologies to analyze single particle trajectories involving either regression fitting of the MSD versus time interval relation, application of a relative deviation parameter or constructed probability distributions representative of the modes of interest.

The propagator functions in effect model the likelihood of an observed track. What our study includes is how the distribution of the parameters which formulate these models can be used to aid in discrimination of the diffusion processes. In fluorescence microscopy, with typically very short tracks observed, it is generally infeasible to analyze such trajectories without some form of population averaging using conventional techniques. Exceptions are made of course to the occasional long track which is observed, or tracks which appear representative, but a majority of the body of data captured is noisy, and unrepresentative if multiple modes of behaviour are under investigation.

Our study was aimed at being able to discriminate, without population averaging, such molecular-level tracks. Once individual particle trajectories are categorized into different modes of diffusive behaviour, models can be built on how they behave collectively, potentially allowing greater physiological interpretation of the protein mobility characteristics in functional, living cells, and hence to have a greater understanding on their underlying membrane micro- and nanoscale structure in a biologically-relevant context. We have performed a validation across the approximately biological relevant parameters *for the datasets presented.* However, extrapolating these to any real system should come with the caveat that the classifications can only really be used as a guide for the particular set of algorithm parameters and system parameters used.

Ultimately, since the inference scheme is probabilistic there will inevitably be some trajectories which are falsely categorized with the wrong behaviour, most often into simple Brownian motion, as shown in the classification matrices. We included details on how model ranking varies with respect to the number of data points to demonstrate that there will often be a crossover between mis-categorization and the correct identification. A caveat then, for interpreting any model selection on the experimental data would be that there is no evidence of heterogeneity *under the given experimental conditions.* If this crossover is unreachable in the experimental framework it will at least inform the experimentalist on the typical minimum duration of track that needs to be detected to permit reliable discrimination (perhaps thereby directing them to change the characteristics of the optical setup and/or the related biological and physico-chemical conditions, such as the type of fluorophore used and whether the application of anti-bleaching reagents is required).

In an earlier single particle tracking study on the Tat system using non-TIRF illumination, the presence of an "immobile" sub-population of TatA protein complexes was reported, but not investigated further [11]. In our study, BARD analysis reveals that a significant proportion of TatA-YFP complex tracks have a confinement radius of 60 ± 40 nm. The measured localization precision on our microscope for tracking a single YFP molecule is ~40 nm. However, TatA complexes were observed to have a broad range of stoichiometry, with a median value of equivalent to ~20-30 TatA-YFP subunits, consistent with that reported previously [11]. These complexes are therefore brighter than a single YFP molecule by a factor of ~20-30, with the localization precision following iterative Gaussian fitting of the intensity profile of these fluorescent spots scaling approximately by the square-root of this factor, or ~5 (see ref. [57]), so the localization tracking precision for most TatA-YFP complexes is more like 5-10 nm. Therefore, the estimated confinement radius here is substantially higher than the localization precision for diffusing complexes, which strongly suggests that the majority of the "immobile" TatA complexes previously reported were in fact exhibiting true *confined* diffusion.

Similarly, we observed a significant sub-population of tracks for the Helix1021-GFP strain which exhibited confined diffusion, here with a mean confinement radius of 110 ± 50 nm, within experimental error of that measured for the TatA-YFP strain. The fact that the transmembrane helix probe has no known specific interaction with molecular systems in *E. coli* suggests that the confinement domains in both cell strains may be represent an intrinsic feature of the cell membrane itself. Similar size putative confinement domains were observed previously in single particle tracking studies of an unrelated bacterial oxidative phosphorylation (OXPHOS) membrane protein [28]. This behaviour had been previously attributed to a possible "respirazone" effect [49] in which different OXPHOS enzymes were pooled together into the same confinement domains to improve electron transfer efficiency throughout the OXPHOS system. However, our work here may point to a more generic confining feature of the cell membrane.

The diffusion models illustrated here are not exclusive as such – there is a risk that none of the models is actually the physically "correct" one. BARD analysis will provide probabilistic rankings of these models, but these probabilities can strictly only be interpreted in the context of the other models considered, and do not represent an absolute probability. Model selection is open-ended; the models presented here do not take into account the full degree of potential heterogeneity that may exist in the cell membrane, and other models can be considered. For example, there are several theorised models of anomalous sub-diffusion, each with a unique PDF. There may also be complex dynamic behaviour that has not been taken into account, such as hopping diffusion, reaction kinetics and molecular assembly effects. A natural extension of this BARD approach as we present it here is to incorporate more complex behaviour which may better capture the real, physiological behaviour of diffusion in living cells.

Separating different mobility characteristics into different categories will clearly facilitate insight into several important biological questions. For example, how proteins partition dynamically in the cell membrane, whether signalling events are linked to membrane architecture, the precise manner in which motor proteins shuttle in or near to cell membranes, and the extent to which interacting proteins rely upon random collisions or are part of putative confined "solid-state" reaction zones. Such new diffusion analysis tools that we report here might indeed also be further extended to larger length scale investigations beyond that of the single molecule and single cell, such as rheological or cell migration studies at the level of cellular populations in normal tissue development and tumour formation in cancer.

**SUPPORTING MATERIAL**

Electronic supplementary material is available via http://rstb.royalsocietypublishing.org.

**ACKNOWLEDGEMENTS**

The authors thank Philip Maini and Marcus Tyndall for preliminary discussions concerning diffusion simulation; Nick Greene and Ben Berks for the donation of bacterial cell strain AyBC, Anja Neninger and Conrad Mullineaux for the donation of bacterial cell strain Helix1021-GFP.  This work was supported via a research grant to MCL (EP/G061009). MLC was supported by a Royal Society University Research Fellowship. AR was supported by the Research Councils UK.

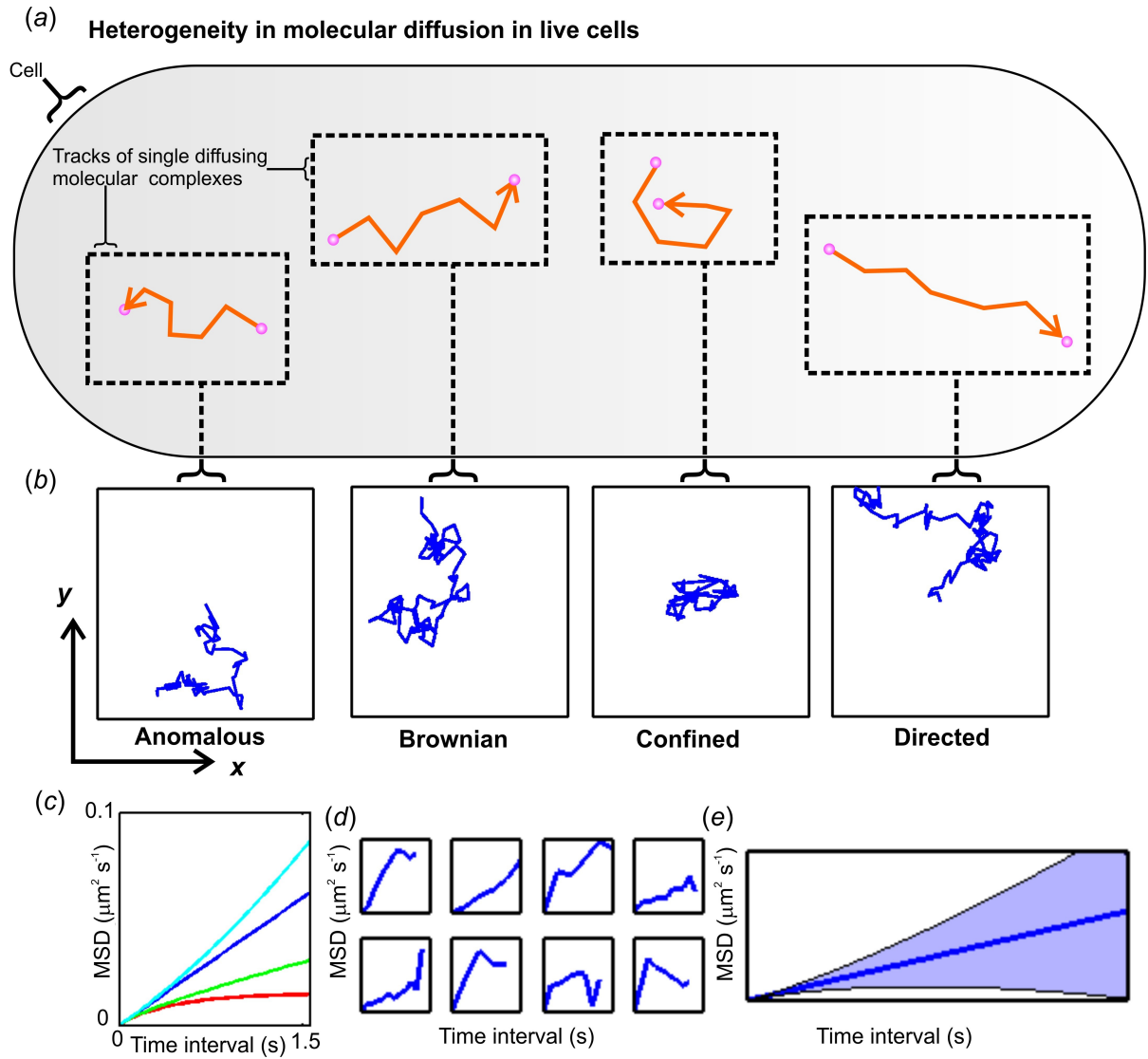

Figure 1. The complexity of molecular diffusion in living cells. (*a*), Schematic of diffusion (orange track) of a single molecule or complex (pink circle) in a live cell with (*b*) simulated data for the typical spatial localizations of tracked molecules in the specimen *xy* plane of a typical optical microscope illustrating four common modes of diffusion. (c) Idealized forms of mean square displacement (MSD) relations for anomalous (*green*), Brownian (*blue*), confined (*red*) and directed (*cyan*) diffusion. (*d*) Typical simulated Brownian diffusion MSD data for different diffusing molecules modelled with the same diffusion coefficient and (*e*) the mean average of many simulated MSD relations (*blue* line), blue shading indicating the stochastic spread at one s,d. illustrating the marked heterogeneity due just to molecular stochasticity.

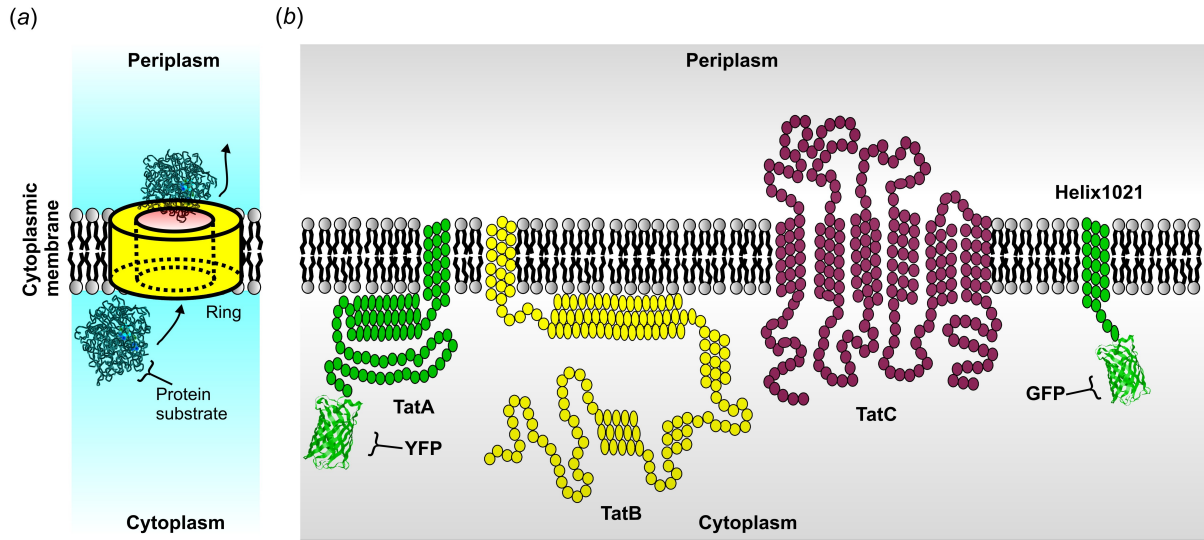

Figure 2. Schematic of (*a*) membrane bound architecture of a Twin-Arginine Translocase TatA nanopore allowing translocation of a full folded protein substrate across the cytoplasmic membrane, and of (*b*) the TatA-YFP, TatB, TatC and Helix1021-GFP proteins (coloured circles indicating individually amino acid residues in the native proteins).

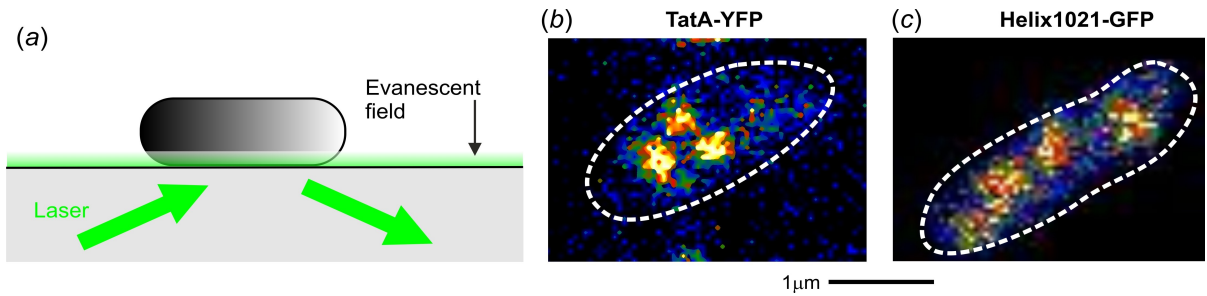

Figure 3. *In vivo* TIRF microscopy on live bacteria. (*a*) Schematic of TIRF on an *E. Coli* cell immobilized to a glass microscope coverslip. False-colour TIRF images on single cells for the (*b*) TatA-YFP and (*c*) Heli1021-GFP cell strains, cell outline indicated (dashed lines).

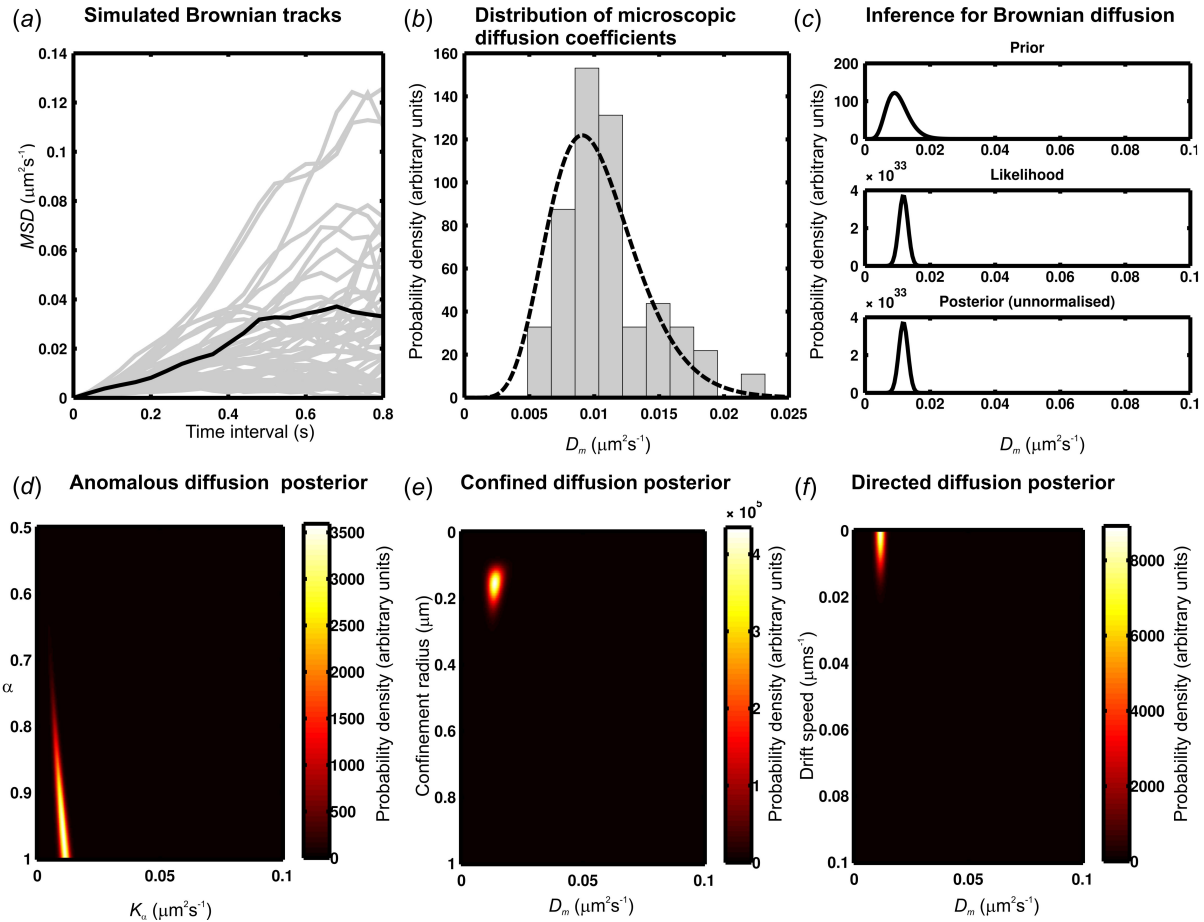

Figure 4. Implementing BARD. (*a*) Simulated Brownian tracks (grey) all with diffusion coefficient $D$=0.01 $\mu m^2 s^{-1}$, with one of these tracks highlighted (black) for BARD analysis. (*b*) Distribution of measured microscopic diffusion coefficient $D_m$ values from tracks in (*a*) with Gamma fit indicated (dashed line). (*c*) Constructing probability distributions used in BARD for the highlighted track of (a) tested against a Brownian diffusion model showing the unnormalized prior (upper panel), likelihood (middle panel) and posterior (lower panel). Testing against the three other diffusion model generates two-dimensional unnormalized posterior distributions for (*d*) anomalous, (*e*) confined and (*f*) directed diffusion models. For the highlighted track shown in (*a*) the highest ranking probability was measured at 65% for the Brownian mode, thus correctly identified with a probability which was more than twice as much as the next ranked mode of anomalous diffusion.

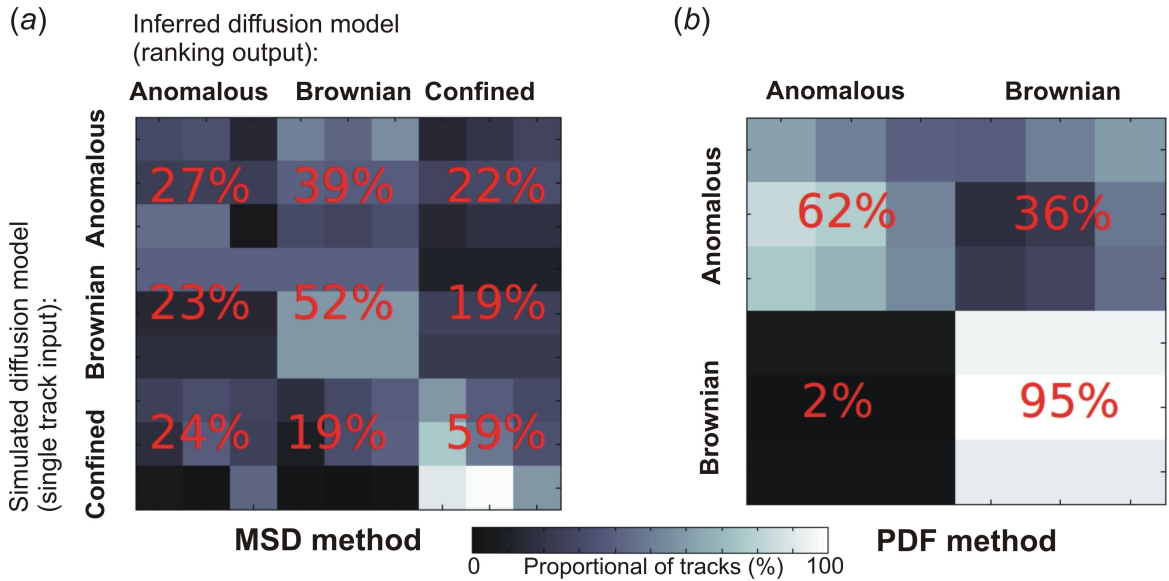

Figure 5. Characterizing the error in the BARD analysis using classification matrices for (*a*) the MSD method, and (*b*) the PDF method. Each "pixel" in the matrix represents a combination of a type of simulated track against an inferred highest ranked diffusion model from the BARD analysis, with greyscale indicating the individual inference ranking percentage. Each diffusion mode here was simulated with three different tracks, and therefore each matrix is composed of individual 3x3 sub-matrices, with the percentage in red indicating the mean average inference ranking percentage within that sub-matrix. Thus, the "diagonal" of (*a*) and (*b*) constitutes the ""true-positive" classification values.

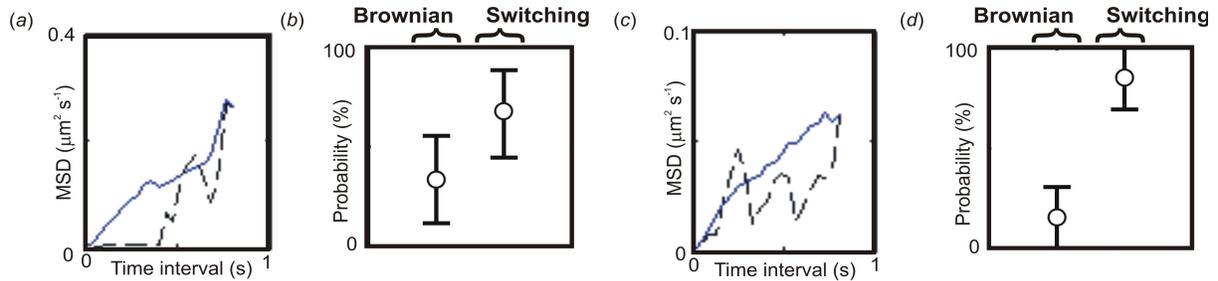

Figure 6. Inferring a sudden change in diffusion coefficient. (*a*) MSD relation for a single track simulated assuming Brownian diffusion (*black*) and the mean average of 20 such tracks (*blue*) for which the diffusion coefficient $D$ switches from 0.01 to 0.10 $\mu m^2\ s^{-1}$, with (*b*) associated ranking inference probabilities for a single D Brownian (B) and two D Switching (S) model. (*c*) MSD relation for a single track (*black*) and average of several tracks (*blue*) for which $D$ switches from 0.01 to 0.05 $\mu m^2\ s^{-1}$, with (*d*) associated ranking inference predictions.

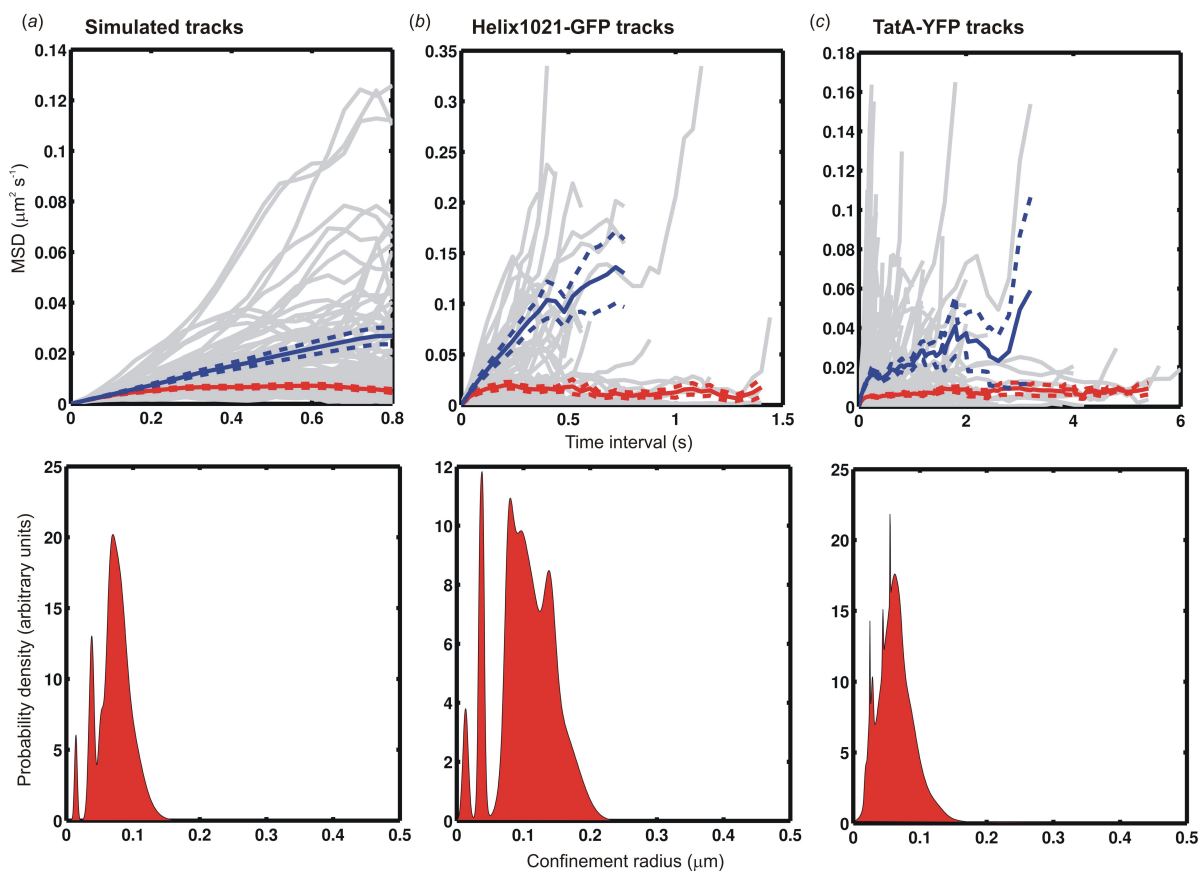

Figure 7. Comparing mobile and immobile tracks. Upper panels are MSD versus time interval traces showing mean values assuming at least *n*=3 data points at each time interval value, for all "mobile" tracks (blue, solid line) and confined tracks (red, solid line), s.e.m. error bounds shown (dotted lines), individual tracks shown in grey, for (*a*) simulated tracks (*n*=50 mobile tracks here simulated using a Brownian diffusion propagator function, *n*=50 confined tracks), (*b*) Helix1021-GFP (*n*=20 mobile tracks, *n*=27 confined tracks) and (*c*) TatA-YFP (*n*=258 mobile tracks, *n*=164 confined tracks). Lower panels are corresponding unbiased kernel density estimations for the distribution of predicted confinement radius from the inferred confined tracks.

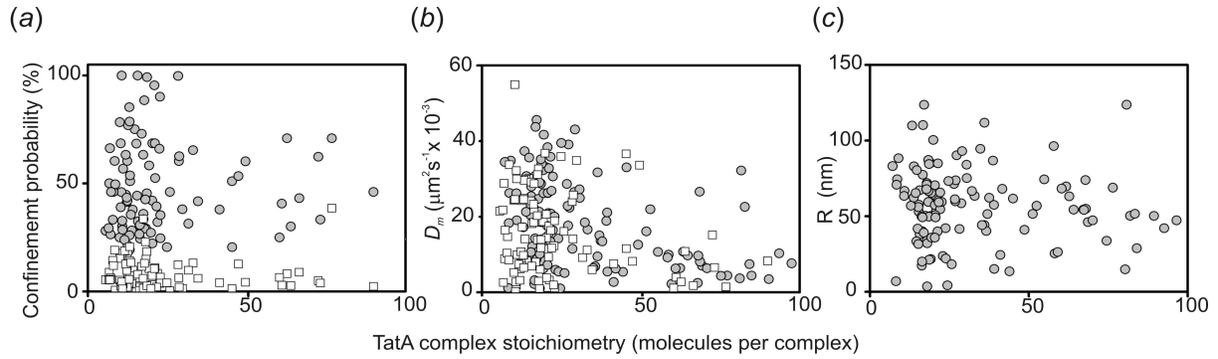

Figure 8. Variation in (*a*) the probability that the top ranked diffusion mode will be confined, (*b*) the microscopic diffusion coefficient $D_m$ and (*c*) the confinement radius $R$ versus estimated molecular stoichiometry for the TatA-YFP complexes, predicted confined (white squares) and mobile (grey circles) tracks indicated.

# Supporting Material

# Inferring diffusion in single live cells at the single molecule level

Alex Robson, Kevin Burrage, and Mark C. Leake*

*For correspondence email: m.leake1@physics.ox.ac.uk

*Generation of synthetic tracks*

2D simulated tracks were generated using a stochastic random walk process, generated in MATLAB (The MathWorks, Natick, MA). This algorithm modelled the stochastic Ito differential equation:

$$dX = F(X,t)dt + \sqrt{2D}\,G(X,t)dW(t) \quad . \tag{S1}$$

Here, $X$ is simulated particle displacement for a given spatial dimension as a function of time $t$, $F$ is a function that represents drift, $G$ is a function that represents diffusion and $dW$ represents an incremental Wiener process. Following an incremental sampling time $\delta t$, $X$ changes its amount by a value $\delta X$ which is normally distributed such that $<\delta X> \approx F(X,t)\delta t$ and the variance $\text{Var}(\delta X) \approx 2DG(X,t)^T G(X,t)\delta t$, where $D$ is the corresponding lateral diffusion coefficient. For Brownian, confined and directed diffusion, appropriate conditions on $F$ and $G$ were chosen (Table S1) using a video-rate sampling time of 40 ms throughout to predict random incremental displacements for a simulated particle track. Anomalous motion was simulated separately, by rejection sampling of the probability distribution but used essentially the same randomized, incremental process. For confined tracks, the domain was modelled as an harmonic potential well in which a particle at the domain edge experiences a forcing function $F$ that drives it back into the domain, similar to the approach in [58] (Figure S1). We simulated anomalous diffusion through a rejection sampling algorithm.

**Table S1. Selection of drift and diffusion functions in the stochastic equation**

| Mode | F(X,t) | G(X,t) |
|---|---|---|
| **Brownian** | 0 | 1 |
| **Directed** | >0 | 1 |
| **Confined** | -sign($X$) if $X \geq L$; 0 if $X < L$ | 1 |

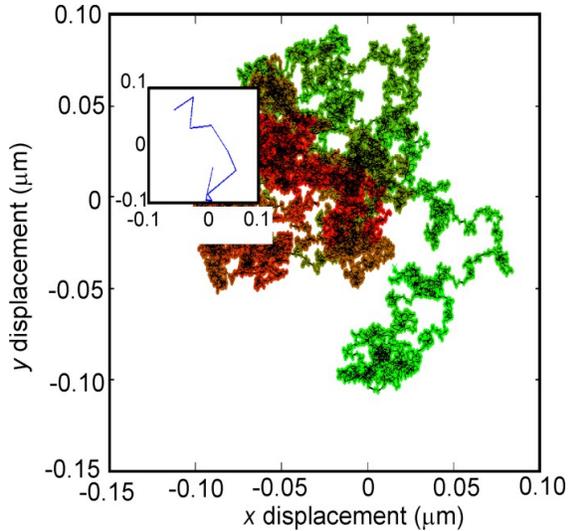

Figure S1. **Simulated confined diffusion**. **Displacements in *xy* of a highly sampled confined track. Colouring added to indicate procession with time proceeding from red at the origin to light green. Inset shows the observed track with just *N* = 10 data points at video-rate sampling. Confinement radius *R* = 0.1 μm.**

*Bayesian formulation*

The general principle of Bayesian inference is to quantify the present state of knowledge and refine this on the basis of new data, underpinned by *Bayes' theorem*, emerging from the definition of *conditional* probabilities. This can be explained by considering the probability of two general events, *A* and *B*, happening, which is denoted P(A∩B), which equates to the probability of *B* happening, P(*B*), multiplied by the probability of *A* given that *B* has occurred, denoted P(*A*│*B*), or:

P(A∩B)=P(*A*│*B*)P(*B*)

Using the same notation we can say that:

P(B∩A)=P(*B*│*A*)P(*A*)

Since these two equations are equal this leads to Bayes' theorem of:

P(*A*│*B*)= P(*B*│*A*)P(*A*)/P(*B*)                                                  (S1)

There are two principle levels of inference [51]. The first is at the level of an individual model, *M*, evaluating the likelihood of the data *d*, P(*d*|*w*,*M*), using the

appropriate function that describes the probability distribution governing the mode. This posterior distribution also incorporates any prior understanding on the set of parameters *w* that comprise that model. Such priors embody our initial estimate of the system, such as the expected order of magnitude or distribution of the parameters. The prior probability distribution, *P*(*w*|*M*), is independent of the data of the system. The results of this inference are summarized by the most probable parameter values and their associated distributions, embodied in the posterior distribution of the parameter as:

$$P(w|d,M) = \frac{P(d|w,M)P(w|M)}{P(d|M)} \quad . \tag{S2}$$

After this stage, model comparison takes place in which models are ranked, conditioned by the observed data to assign a probability-based preference between the distinct models:

$$P(M|d) = \frac{P(d|M)P(M)}{P(d)} \quad . \tag{S3}$$

Provided that the model priors are "flat" (i.e. no particular *a priori* preference, which is our default assumption in all diffusion models with the exception of confined diffusion), the result of model evaluation is by simply ranking the *marginal likelihood*, *P*(*d*|$M_i$), of each individual mode $M_i$. This is also known as the evidence *E* which is given by the denominator in equation S2 by integrating the data over the entire parameter space:

$$E = P(d|M) = \int P(w|M)dw \quad . \tag{S4}$$

The final probability that a given model *M* results in data *d* for any given single track is then given by:

$$P(M|d) = \frac{P(d|M)P(M)}{\sum_i P(M_i)P(d_i|M_i)} \quad . \tag{S5}$$

Put more simply in words, the product of the calculated likelihood of a specific diffusion model, given the (*x,y,t*) localization data from a single track, with any relevant parameter priors is proportional to the posterior distribution. The normalisation quantity in this, equation S3 calculated by equation S4, is then used in equation S5. This forms the model likelihood, and with another application of Bayes Theorem, equation S1, using a model prior. We then calculate the model probability. This is the quantity used to rank models.

**Integration routine**

The evidence/normalisation term is numerically calculated by a Monte Carlo integrator. The integration routine used a Monte Carlo approach derived from
http://www.mathworks.com/matlabcentral/fileexchange/12447-mcint
(Lee Ferchoff on Mathworks.com MATLAB Central, *unlicensed*). This is not a regular uniform sampling but a Monte Carlo approach, 200 sampling points were used in this.

*Diffusion Models*

In the first instance, four diffusive modes were modelled, chosen to be typical of a variety of different biological phenomena which can be observed at the single molecule level by single particle tracking, namely normal Brownian diffusion, anomalous or sub-diffusion, confined diffusion and directed diffusion. Brownian and anomalous diffusion represent different solutions of the governing diffusion equation and the fractional diffusion equation respectively. However, because the diffusion process can be subject to imposed boundary conditions, and an advective component introduced, they may also be solved to reflect confinement effects as well as the effects of directed diffusion.

For a particle initially located at spatial point *x*, at time *t*, which is found at *x'* at the later time $t' = t + \delta t$, where $\delta t$ is the sampling period, the two-dimensional diffusion-(advection) equation is given by:

$$\frac{\partial W}{\partial t} + v \nabla (W) = K_1 \nabla^2 W(x'-x, t) \quad . \tag{S6}$$

The fractional version, for anomalous sub-diffusion, is:

$$\frac{\partial W}{\partial t} = \left[ -v \nabla + {}_0 D_t^{1-\alpha} K_\alpha \nabla^2 \right] W(x'-x, t) \quad . \tag{S7}$$

Here, **v** is the velocity flow vector in the case of directed diffusion, with corresponding speed *v*. Equation S7 introduces the fractional Riemann-Liouville operator $_0D_t^{1-\alpha}$ (see ref [34]). It reduces to equation S6 when $\alpha = 1$. Equations S6 and S7 can be solved, subject to any imposed physical boundary conditions, to generate the probability density distribution function *W*, commonly referred to as the *propagator.* The full list of propagators investigated here are given in equations S16–S18, with the range of parameters summarized in Tables S2 and S3, prior estimates based on real experimental data both from this study and from earlier cited investigations.

**Table S2. The propagators (2D form) and MSD functions used for the four diffusive modes**

| Mode | Propagator |
| --- | --- |
| **Brownian/Directed** | Normal diffusion propagator solution [12] |
| **Confined** | Series solution to Equation S6 reflecting boundary conditions [26] |
| **Anomalous** | Numerical solution for time-space fractional diffusion [59] |

**Table S3. Typical mean equation parameters used in diffusion propagator functions, with cited sources for range of values**

|  | Diffusion Coefficient | Domain radius | Transport coefficient | Anomalous exponent | Flow speed | Sampling rate |
| --- | --- | --- | --- | --- | --- | --- |
| **Reference Sources** | $D=0.01$ μm$^2$s$^{-1}$ <br> This work and [5,11, 27, 28] | $R=0.1$ μm <br> [4, 9, 16, 26, 27] | $K_\alpha=0.01$ μm$^2$s$^{-1}$ <br> [33-35] | $\alpha=0.75$ <br> [33-35] | $v=10$ nm s$^{-1}$ <br> [28, 29] | $1/\delta t=25$ Hz |

There is an important cautionary note for those investigating the actual sources of anomalous diffusion. Anomalous behaviour can typically arise through either "transient trapping" effects or through "molecular crowding" of the environment. Three types of anomalous diffusion propagator functions have been considered in recent literature - fractional Brownian motion, percolation propagator functions and continuous time random walks, with the latter creating non-*ergodic* behaviour. Heterogeneous SPT data therefore can potentially be ascribed to three different sources of variation; the unavoidable statistical spread due to sampling a finite time-series, cell membrane structural/chemical heterogeneity of the protein environment, and potentially non-ergodic diffusive processes - which means the very method of averaging (whether using "ensemble averaging", i.e. averaging over a lot of different tracks, or "time averaging", i.e. averaging over the same track but over a very long time) has implications.

*Inference scheme*

The inference scheme is split into two forms – one using the probability distributions directly, and the other using the mean square distance distribution. All data *d* comes from the (*x,y,t*) data of each track, although may be used as either the time-averaged MSD or 2-dimensional spatial displacements.

*PDF method.* The likelihood is found by evaluating the probability of each observed particle displacement. Each trajectory is composed of an *N* length vector of two coordinates, **C**, sampled uniformly over time *t*. The probability of each displacement over a time *t*, from the $n^{th}$ to the $(n+1)^{th}$ coordinates, $\mathbf{C}_n = (x_1, y_1)$ to $\mathbf{C}_{n+1} = (x_2, y_2)$, is calculated using the propagator, $W(\mathbf{C}_{n+1}, \mathbf{C}_n, t)$, of each model, *M*, which is parameterized by the set of parameters *w*. The set of coordinates **C** are used to form the displacements being the pair-wise differences of each coordinate. This is over each time window *t*, which is given as multiples of the incremental sampling time $\delta t$. These form the data, *d*. As there are a total of *N* data points, the likelihood is the product of *N*-1 evaluations of each displacement. The actual PDF propagators used are given in the Equations S16-S18. The likelihood is then given by:

$$P(d\mid M,w) = \prod_{n=1}^{N} W\left((C_{n+1}\mid C_n)\mid w, M\right) \quad \text{(S8)}$$

*MSD method.* The likelihood is found by assuming normal distribution of errors about the MSD. The likelihood for a track consisting of *N* data points occurring a time interval values of $t_1, t_2, \ldots, t_N$, is given by:

$$P(d|\theta)=\prod_{i=1}^{N}\frac{1}{\sqrt{2\pi\sigma^{2}(t_{i})}}\exp\left[-\frac{1}{2}\sum_{i=1}^{N}\left(\frac{d(t_{i})-M(\theta,(t_{i}))}{\sigma(t_{i})}\right)^{2}\right] \quad (S9)$$

Here, the data *d* referred to is given by the time-averaged MSD, with *M* the equivalent theoretical prediction for a given diffusion model, and σ is given by:

$$\sigma(t_i)=\frac{d(t_i)}{\sqrt{t_N/t_i}} \quad (S10)$$

The transport coefficients (parameters characterizing mobility) in each case are modelled by a Gamma probability distribution, $\Gamma(k,\theta)$, parameterized in terms of the expected diffusion value $D_m$ and the standard deviation, $\sigma_D$. The Gamma distribution is the natural representation of the diffusion coefficient [56], a positive Gaussian-like distribution around peak values. The Gamma probability density function *f* for a variable *x* parameterized by the shape parameter *k* and scaling parameter *θ*, is

$$f(x,k,\theta)=x^{k-1}\frac{e^{-x/\theta}}{\theta^{k}\Gamma(k)} \quad (S11)$$

Here, *x*, *k* and *θ* are > 0, and $\Gamma$ is the Gamma function. The values $D_m$ and $\sigma_D$ are used to choose the parameters of the Gamma distribution, *k* and *θ*, such that the mean resides at $D_m$ and the standard deviation is equal to $\sigma_D$. The probability density function of the prior is then given by:

$P(d)=f(d,\sigma_D/D_m,D^2_m/\sigma_D).$     (S12)

For the case of anomalous sub-diffusion, we mapped the transport coefficient $K_\alpha$ to *d* by using $d = K_\alpha\Gamma(1 + \alpha)$ (see ref. [35]). In general, appropriate values of the $D_m$ can be estimated from other experimental data, or by heuristic mobility measurements from other tracks in the same dataset, and in our study here we have used several such past estimates which generated estimates for diffusion coefficients of membrane protein complexes of similar molecular weights (see Table S3). The α coefficient for the anomalous diffusion model was assumed to be uniform, characterized by a Heaviside function assumed to be 2 is the range 0.5-1.0 and zero elsewhere. The two other independent parameters in the standard diffusion models investigated here of domain radius *R* for confined diffusion and the particle mean drift speed *v* for directed diffusion were *regularized* by an appropriate smoothing function such that they had an exponential decay function. [52] of the form:

$$\langle w \rangle = \int w \left( \frac{2\exp(-\lambda^2 w^2)}{\sqrt{\pi/\lambda^2}} \right) dw \quad . \tag{S13}$$

Here, $\lambda$ is a parameter with a value chosen to ensure the bulk of the probability mass was within an appropriate range such that $1/\lambda$ is equal to <*w*>, approximated by using physically realistic assumptions to generate mean values characteristic of those found in previous experimental studies (Table S3). The parameter $\lambda$ is a so-called *hyperparameter*. It can be *marginalized* (integrated out), but given that the scales of the system do not change from track to track, this can be approximated by using physically realistic assumptions. The effect of $\lambda$ is investigated via a sensitivity analysis.

To explore the dependence of the correct inference probability for confined diffusion on the shape of the prior distribution we simulated confined diffusion using the expected parameters values of Table S3, and then performed BARD analysis assuming the standard four candidate diffusion models of anomalous, Brownian, confined and directed, and then measured the proportion of tracks that were correctly inferred as confined as a function of the confinement radius *R* which was used in the prior function, in the range 50-200 nm.

The correct probability of inference was found to have a mean of ~61% but to only vary by a few % across the range of different *R* values. In other words, the inference probability is relatively insensitive to *R* across a large range of physically realistic *R* values that differ by a factor of ~4. This is not to say that the prior function has little effect on the final outcome; rather, what the value of $1/\lambda$ does here is to set an inference penalty for values of *R* that are too high to be physically realistic – for example, that are larger than the length scale of a single cell. We can sees this when we run BARD analysis on the same track dataset but assuming a "flat" prior function for confined diffusion (in effect, taking a value of $R \to \infty$), in which case the probability of inference of confined diffusion is only 50%. This means that the effect of using a sensible prior function here increases the proportion of correctly inferred confined tracks by over 20%.

Conversely, to correctly infer confined diffusion from a given single track requires that a typical confined particle has actually diffused for a large enough time to allow it to experience the confining boundaries, otherwise it will just appear to exhibit Brownian diffusion, which means there is an expected dependence on the number of data points for the correct ranking inference for confined diffusion. To explore this effect we simulated confined tracks again with characteristic typical parameters of earlier studies (Table S3), but varied the number of data points in each track from 2-40, and ran a BARD analysis using just two candidate diffusion models of Brownian and confined diffusion. This indicated that tracks have a probability of more than 50% of being correctly inferred as being confined if they contained greater than ~16 data points (figure S2).

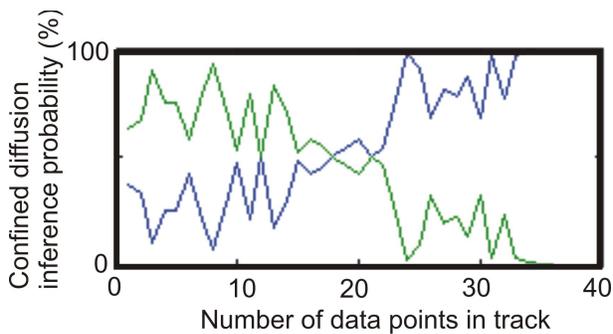

**Figure S2.** Variation of ranking probability with number of track data points. A single track was simulated using a confinement radius 0.1 μm and microscopic diffusion 0.01 μm$^2$ s$^{-1}$. Model ranking inference was then applied against two possible diffusion models of confined (*blue*) and Brownian (*green*) diffusion, with the vertical axis here indicating the inference ranking probability for confined diffusion.

When performing a BARD analysis on simulated directed diffusion tracks using transport parameters appropriate to an earlier treadmilling study [29], and then using two candidate inference models of Brownian and directed diffusion, the proportional of tracks that were correctly inferred as directed diffusion was 60-70% (figure S3).

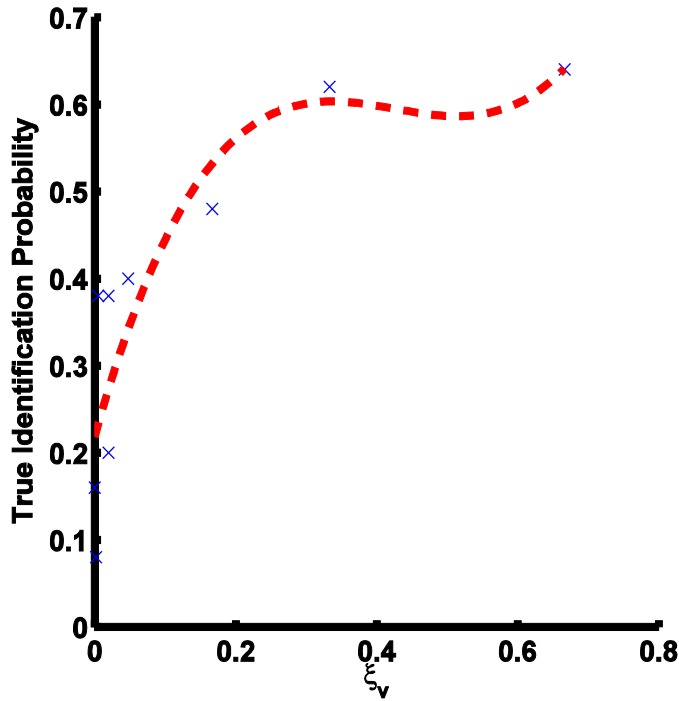

**Figure S3.** Directed diffusion validation. A two-dimensional single track was simulated assuming confined diffusion using confinement radius 0.1 μm and microscopic diffusion 0.01 μm² s⁻¹, and model ranking inference was then applied against two possible diffusion models of confined (*blue*) and Brownian (*green*) diffusion, with the vertical axis here indicating the inference ranking probability for confined diffusion. Y-intercept and control simulations (where there is no directed motion) gives an estimate of the systematic error at around 20%. $\xi_v$ is a measure of degree of directed to random drift and is given by $\xi_v=(vt)^2/((vt)^2+4Dt)$. The diffusion coefficients simulated were $1\times10^{-5}, 1\times10^{-4}$ and $1\times10^{-3}$ μm² s⁻¹ with velocities of 1, 10 and 20 nm s⁻¹. In addition, three controls were done at the listed diffusion coefficients at zero velocity.

Inferred values for $D_m$, $K_\alpha$, $R$ and $v$ parameters were all estimated from summary statistics of the relevant posterior distribution for a given single track. To summarize the distributions into representative values, a Gaussian was fitted about the posterior maximum.

*Choosing the Priors*

*Parameter priors*

Single particle tracking data and simulations show that the diffusion coefficient tends to be distributed by a Gamma distribution. This is unsurprising, as it represents the statistical spread in the diffusion coefficient for observations of a Gaussian distributed random walk. The mean and standard deviation width of the microscopic diffusion coefficient, $D_m$ and $\sigma_D$ respectively, were calculated by best-fit of the distribution of measured diffusion values. The characteristic domain radius, $R$, represents the typical domain size expected. More strictly, it is the parameter governing the prior on the distribution of domain size. Note that the uniform distribution is recovered as $R\to\infty$. Similar arguments apply to the characteristic velocity.

*Model priors*

The natural choice for model priors are unbiased weights. However, with confined diffusion, there is a clear difference between long and short tracks. With longer tracks, diffusing molecules and complexes may explore the entire domain exhibiting a corralled motion, but this will be limited with short tracks. We quantify this by using a corrective prior on the confinement, $P(C)$ given by:

$$P(C) = P'(C)\left[1-\exp\left(-D_m t_N / \lambda^2\right)\right] \quad . \tag{S14}$$

where $\lambda$ is the characteristic domain size and $P'(C)$ is the probability of tracks we expect to be confined, independent of any parameters. Irrespective of the number of confinement zones, short tracks cannot probe domains like long tracks – we do not expect to see confinement. This does not extend to directed drift or sub-diffusive behaviours. As there is no prior expectation on any of the diffusion models, $P'(M)$ is assumed to be the same for each.

**Diffusion propagator functions**

The Riemann-Liouville Operator, $_0D_t^{1-\alpha}$ is defined as:

$$_0D_t^{1-\alpha}W(x,t) = \frac{1}{\Gamma(\alpha)}\frac{\partial}{\partial t}\int_0^t \frac{W(x,t')}{(t-t')^{1-\alpha}}dt' \quad . \tag{S15}$$

Here, Γ is the Gamma function, $W$ the diffusion propagator function. Anomalous motion is given by a complicated propagator, of which we use numerical approximations, but the full solutions can be found in ref. [34]. The theoretical distributions governing the transport behaviour for the different diffusive modes applied here, used to evaluate the likelihoods, are given in the two-dimensional probability density functions below for anomalous sub-diffusion, Brownian motion and directed diffusion. A confinement probability distribution was not used here, because this requires using the absolute spatial positions over the relative difference. This would involve marginalizing over the start point and resulted in weak performance or infeasible computational load.

Anomalous:

$$W(\mathbf{r},t) = \sum_{m}^{\infty} \sum_{n}^{\infty} E_{\alpha,1}(-\lambda_{mn} K_{\alpha} t^{\alpha}) \delta_{mn} \beta_{mn} \quad . \tag{S16}$$

Here, $\delta_{mn}$ is the Kronecker delta function, $\beta_{mn}$ are the Fourier coefficients of the initial condition defined where $W(r,0)$ is a delta function, and $E$ is the Mittag-Leffler function, which has a general definition of:

$$E_{a,b}(z) = \sum_{k=0}^{\infty} \frac{z^k}{\Gamma(a+kb)} \quad . \tag{S17}$$

Brownian and directed:

$$W(\mathbf{r},t) = \frac{1}{4\pi Dt} \exp\left( \frac{1 |\mathbf{r}+\mathbf{v}t|^2}{4Dt} \div \right) \quad . \tag{S18}$$

Here, **r** and **v** are the positional and velocity vectors respectively, with $D$ the diffusion coefficient.

Other examples of priors which could be applied in modification of our scheme include the Einstein-Stokes model of $D = k_B T/\gamma$, where $\gamma$ is the frictional drag coefficient, $k_B$ is the Boltzmann constant and $T$ is the absolute temperature. This could be done by using the heuristic $1/r$ dependence on $\gamma$ where $r$ is the effective membrane protein radius. This was the model used when applied to our in vivo experimental data for the Tat bacterial system. As a general rule, our scheme can be extended to use any general prior on $D$ that can be constructed from the available structural information of the protein and membrane, thermodynamic quantities, and the geometry of the protein in the membrane.

**MSD analysis**

The MSD function for a track of $N$ consecutive image frames at a time interval $\tau = n\Delta t$ was defined according to ref. [25] for a track in 2-dimensional space assuming a standard orthogonal $xy$ coordinate system, given in the following:

$$MSD(\tau) = MSD(n\Delta t) = \frac{1}{N-1-n} \sum_{i=1}^{N-1-n} \left\{ [x'(i\Delta t + n\Delta t) - x'(i\Delta t)]^2 + [y'(i\Delta t + n\Delta t) - y'(i\Delta t)]^2 \right\}.$$

(S19)

A small modification was used when compiling MSD data from experimental sources in that the MSD at the $\tau = 0$ point was defined as $2\sigma^2$ where $\sigma$ is the localization precision for a given detected spot of intensity [55] set at 40nm.

Anomalous diffusion (see ref. [31]):

$MSD = K(K'_\alpha t^\alpha / \Gamma(1+\alpha))$ (S20)

Here, the anomalous transport parameter $K$ is equivalent to the diffusion coefficient $D$ of Brownian diffusion. In order to separate the dependency of $K$ on $\alpha$, we decouple by setting $K$ with units of m²/s and defining $K'_\alpha = 1$ second$^{1-\alpha}$.

Brownian and directed diffusion (see ref. [25]):

$MSD = 4Dt + (vt)^2$ (S21)

Here, $v$ is the mean drift speed and is set to zero for Brownian diffusion.

Confined diffusion (see refs. [26] and [55]):

$$MSD = R^2 \left(1 - 8\sum_{m=1}^{\infty} \exp\left[\beta_m^2 \frac{t}{(R^2/D)} \frac{1}{\beta_m^2(\beta_m^2-1)}\right]\right) \quad (S22)$$

Here we modelled confined diffusion in a circular domain of radius $R$. $\beta_m$ is equal to the $m^{th}$ root of the first order Bessel function $J(\beta_m)$.

**Switching**

To identify "switching" a routine similar to that presented in the main text was used. This introduced a new model with Brownian diffusion before and after a switching event but in which the diffusion coefficient changed. This was with two Gaussians (Brownian propagators) with two independent diffusion coefficients. The diffusion prior was estimated from an ensemble of such tracks. As this was proof of principle, the point of switching was known. To break the dependency on this assumption, a more accurate approach would have a third parameter in which this transition point was inferred.

**Sensitivity Analysis**

In table S4 the results of the sensitivity analysis for confinement model inference are shown. Two test datasets were used, one of free diffusion and the other of diffusion in confinement. The increase in model sensitivity is shown by using an appropriate prior over the confinement size. Although a 'flat' prior has the greatest sensitivity to identify confinement, this decreases sensitivity when identifying free diffusion. When calculating sensitivity by account for both test data sets a finite $\lambda$ (see equation S13) has the most even sensitivity between for both models. Consequently, a non-flat prior is preferred, and the regularised Gaussian decay is used. Overall, when viewing both model selection sensitivity and inferred parameter accuracy (not shown); a finite value of $\lambda$ has unskewed model selection and appropriate parameter inference, giving a small advantage over a uniform prior. The integration routine here used different boundary conditions due to the wider variance in parameters; however the boundaries were chosen such that the sampling density was the same across the probability peak.

Confined: $R=100$ μm, $D=0.01$ μm$^2$ s$^{-1}$, number of tracks $n=50$

| Mode | Characteristic Prior Parameter ($1/\lambda$, see equation S13) | | | |
|---|---|---|---|---|
| | Flat | 0.2 | 0.1 | 0.05 |
| A | 20 | 22 | 26 | <span style="color:red">32</span> |
| B | <span style="color:green">14</span> | 26 | <span style="color:green">24</span> | 14 |
| C | <span style="color:green">62</span> | 40 | <span style="color:green">44</span> | <span style="color:green">40</span> |
| D | 4 | 12 | 6 | 14 |

Brownian: $D=0.01$ μm$^2$ s$^{-1}$, n=50

| | Flat | 0.2 | 0.1 | 0.05 |
|---|---|---|---|---|
| A | 10 | 26 | 12 | 16 |
| B | <span style="color:red">32</span> | 42 | <span style="color:green">46</span> | <span style="color:green">54</span> |
| C | <span style="color:red">42</span> | 22 | <span style="color:red">28</span> | <span style="color:red">12</span> |
| D | 16 | 10 | 14 | 18 |
| True-positive | 47 (31+16) | 41 (20+21) | <span style="color:green">45 (22+23)</span> | 47 (20+27) |

**Table S4. Sensitivity Analysis for confinement inference (values in %)** Upper table are tests on a set of confined tracks with varying parameter prior parameters. Numbers indicate percentage identification of the respective mode (A=Anomalous, B = Brownian, C=Confined, D= Directed) For example, for a parameter ($1/\lambda$) of 0.05, 32% of the tracks were identified as anomalous. Where numbers are highlighted green indicate the most ideal results (upper table, high C percentages, lower table, high B percentages). Red highlights indicate poor results: (upper table: high B per percentages, lower table, high C percentages). The last row indicate the sensitivity accounting for both datasets. The numbers in brackets indicate the summation over the number of true positive classifications. I.e. 45% is 22 correct C identifications out of 50 plus 23 correct B identifications out of 50.

**Supplementary references:**